\documentclass[twocolumn,pra,unsortedaddress,showpacs,floatfix,citeautoscript,nofootinbib]{revtex4-1}
\usepackage{amssymb}
\usepackage{amsbsy}
\usepackage{latexsym,epsfig,graphicx}
\usepackage{dcolumn}
\usepackage{bm}
\usepackage{graphicx}
\usepackage{subfigure}
\usepackage{color}
\usepackage[colorlinks,urlcolor=blue,citecolor=blue]{hyperref}

\begin{document}

\title{Spin--orbital-angular-momentum coupling in Bose-Einstein condensates}
\author{Kuei Sun}
\author{Chunlei Qu}
\author{Chuanwei Zhang}
\thanks{Corresponding author: chuanwei.zhang@utdallas.edu}

\affiliation{Department of Physics, The University of Texas at
Dallas, Richardson, Texas 75080-3021, USA}

\begin{abstract}
Spin-orbit coupling (SOC) plays a crucial role in many branches of
physics. In this context, the recent experimental realization of
the coupling between spin and linear momentum of ultracold atoms
opens a completely new avenue for exploring new spin-related
superfluid physics. Here we propose that another important and
fundamental SOC, the coupling between spin and orbital angular
momentum (SOAM), can be implemented for ultracold atoms using
higher-order Laguerre-Gaussian laser beams to induce Raman
coupling between two hyperfine spin states of atoms. We study the
ground-state phase diagrams of SOAM-coupled Bose-Einstein
condensates on a ring trap and explore their applications in
gravitational force detection. Our results may provide the basis
for further investigation of intriguing superfluid physics induced
by SOAM coupling, such as collective excitations.
\end{abstract}

\maketitle

\section{Introduction}\label{sec:introduction}

Spin-orbit coupling (SOC), the interaction between
a particle's spin and orbital degrees of freedom, takes place in
nature in various ways. For a relativistic spinor, its spin
angular momentum naturally couples to the linear momentum under
Lorentz transformation, constituting the key physics in the Dirac
equation~\cite{Dirac1928}. In solid-state systems, the spin and
linear momentum (SLM) coupling (e.g., Rashba~\cite{Rashba} and
Dresselhaus~\cite{Dresselhaus} coupling) is crucial for many
important phenomena such as quantum spin Hall
effects~\cite{Kato2004,Konig2007,Kane2005,Bernevig2006},
topological insulators, and topological
superconductors~\cite{Hasan,Qi}. Recently, a highly tunable SLM
coupling has been realized in cold atom
experiments~\cite{Lin2011,Pan2012,Qu2013,Hamner2014,Olson2014,Ji2014,karina2014,Campbell2015,Wang2012,Cheuk2012,Williams2013}
using Raman coupling between two atomic hyperfine states
\cite{Spielman2009}. These experimental advances have resulted in
an active field of experimental and theoretical
study~\cite{Dalibard2011,Galitski2013,Zhou2013,Goldman2013,Wang2010,Wu2011,Ho2011,Zhang2012,Hu2012,Ozawa2012,Li2012,Wei2013,Xu13,Xu15b,Gong2011,Hu2011,Yu2011,Qu13,Zhang13,Chen13,Xu14a,Lin14,Xu14b,Jiang14,Xu14c,Xu15a,Jiang15}
on the physics of SLM coupled Bose-Einstein condensates (BECs) and
degenerate Fermi gases.

Another ubiquitous SOC in atomic and condensed matter physics is
the coupling between spin and orbital angular momentum (SOAM). In
a hydrogen atom, the electron's orbital movement generates a
magnetic moment that couples to its spin, leading to SOAM coupling
that is responsible for the spectroscopic fine structure. In
solid-state systems, SOAM coupling plays a crucial role for
magnetic properties of materials~\cite{Herman1958,Zutic2004}.
However, the SOAM coupling for ultracold atoms has not been
realized in experiments and the physics of SOAM-coupled BEC and
degenerate Fermi gases has not been well explored.

\begin{figure}[b]
 \centering
\includegraphics[width=8.6cm]{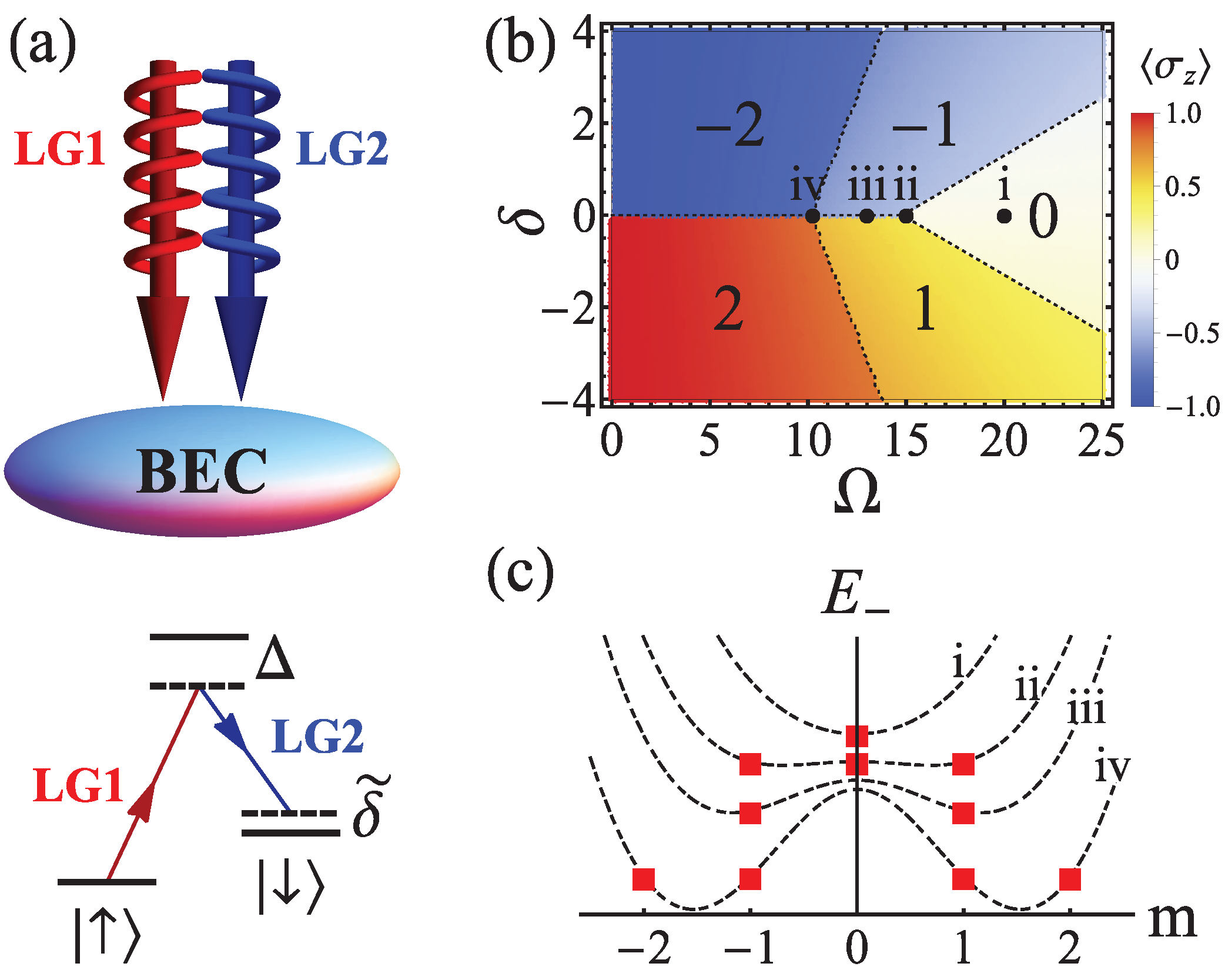}
\caption{(Color online) (a) Illustration showing that two
copropagating LG beams with different OAM couple two internal
states of a BEC through the Raman transition. (b) Noninteracting
ground-state phase diagram for $l=2$ in the plane of detuning
$\protect\delta $ and Raman coupling $\Omega $. The ground-state
OAM quantum numbers are labeled in corresponding blocks separated
by dashed lines. The colors scaled in bar graph represent spin
polarization $\langle \protect\sigma _{z}\rangle $. (c) The ground
states (filled squares) and the assumed continuous spectra (dashed
curves) at selected points in panel (b). (i) Single ground state
$|0\rangle $; (ii) threefold degeneracy $|0\rangle $, $|\pm
1\rangle $ at ${\Omega}_{c}=15$; (iii) twofold degeneracy $|\pm
1\rangle $; and (iv) fourfold degeneracy $|\pm 1\rangle $, $|\pm
2\rangle $. The curves are arbitrarily shifted in $\hat{y}$
direction.} \label{fig:f1}
\end{figure}

In this paper we propose a practical scheme for generating SOAM
coupling for cold atoms and investigate the ground-state
properties of SOAM-coupled BEC. Our main results are the
following:

(1) We propose that the SOAM coupling for cold atoms can be
realized using two co-propagating Laguerre-Gaussian (LG) laser
beams~\cite
{Marzlin1997,Andersen2006,Ryu2007,Leslie2009,Gullo2010,Beattle2013}
that couple two atomic hyperfine states through the two-photon
Raman process~\cite{Juzeliunas2004,Cooper2010} [see
Fig.~\ref{fig:f1}(a)]. Note that only the lowest-order Gaussian
laser beams have been used in the implementation of SLM coupling,
and higher-order LG laser beams are widely available in optical
and atomic experiments. We derive the single-particle Hamiltonian
with SOAM coupling as a function of the laser parameters.

(2) We study the ground-state properties of a SOAM-coupled BEC
trapped on a ring. This geometry has been recently realized in experiments \cite%
{Ramanathan2011,Wright2013,Eckel2014}. We find that the interplay
between SOAM coupling and orbital angular momentum (OAM)
quantization can lead to fourfold degenerate ground states and
first-order transitions between different OAM phases. Both have
not been found in the SLM-coupled BEC. We also find strongly
interacting effects in the system, including a significant
deviation from the single-particle picture and a very large
stripe-phase region.

(3) We show that inhomogeneous potentials, such as gravitational
potentials, can induce the mixture of neighboring OAM states,
leading to the transition from uniform to stripe types of density
distributions. Such a transition may find potential applications
in designing gravitational force detection devices.

The paper is organized as follows. In Sec.~\ref{sec:model} we
derive the model Hamiltonian with SOAM coupling. We then analyze
the single-particle physics of a ring system in
Sec.~\ref{sec:ring} and show interacting phase diagrams for a
realistic ring BEC in Sec.~\ref{sec:interactioin}. In
Sec.~\ref{sec:potential} we study effects of external potentials.
Experimental parameters are discussed in
Sec.~\ref{sec:experiment}. Section \ref{sec:conclusion} is the
Conclusion.

\section{Model and Hamiltonian}\label{sec:model}

As illustrated in Fig.~\ref{fig:f1}(a), we
consider an atomic BEC with two internal spin states, $\left\vert
\uparrow \right\rangle $ and $\left\vert \downarrow \right\rangle
$, coupled by a pair of copropagating Raman lasers. In order to
transfer OAM from the laser to atoms, both Raman lasers are chosen
to be LG beams with different OAM denoted by azimuthal indices
$l_{1,2}$, respectively. The one-photon Rabi frequency from the
$j$-th beam in cylindrical coordinate can be written as
\begin{eqnarray}
\Omega _{j}(\mathbf{r})=\Omega
_{0,j}(\frac{\sqrt{2}r}{w})^{|l_{j}|}\exp \left(
-\frac{r^{2}}{w^{2}}+il_{j}\phi +ik_{z}z\right),
\end{eqnarray}
where $\Omega _{0,j}$ is proportional to the beam intensity, $w$
is the beam waist, $r$ is the radius, and $\phi $ is the azimuthal
angle. Hereafter we consider the case $-l_{1}=l_{2}=l$ for
convenience. The two-photon Raman coupling between two spin states
is $\Omega _{1}\bar{\Omega}_{2}/4\Delta \equiv ({\tilde{
\Omega}}/2)f(r){e^{-2il\phi }}$, with the strength
$\tilde{\Omega}$ and spatial distribution $f(r)$. Incorporating
additional detuning $\tilde{ \delta }/2$, the effective
single-particle Hamiltonian is written as
\begin{eqnarray}
H_{0}= \left( {\
\begin{array}{cc}
{-\frac{\hbar ^{2} \nabla ^{2}}{2M}+\frac{\tilde{\delta}}{2}} &
{\frac{\tilde{\Omega}}{2}f{e^{-2li\phi }}} \\
{\frac{\tilde{\Omega}}{2}f{e^{2li\phi }}} & {-\frac{\hbar ^{2}
\nabla ^{2}}{2M}-\frac{\tilde{ \delta}}{2}}
\end{array}
}\right) +V(r),
\end{eqnarray}
 in basis $\Psi ={\left( {\
\begin{array}{cc}
{\psi _{\uparrow },} & {\psi _{\downarrow }}
\end{array}
}\right) ^{T}}$, where $V(r)=\mathrm{diag.}\left( {\
\begin{array}{cc}
{{{|{{\Omega _{1}}}|}^{2}/4\Delta },} & {{{|{{\Omega
_{2}}}|}^{2}/4\Delta }}
\end{array}
}\right) $ describes the Stark shift~\cite{Marzlin1997} and $M$ is the
atomic mass. After a unitary transformation $\psi _{\uparrow /\downarrow
}\rightarrow e^{\mp il\phi }\psi _{\uparrow /\downarrow }$, we obtain
\begin{eqnarray}
{H_{0}^{\prime }} &=&\frac{{\hbar ^{2}}}{{2M{r^{2}}}}\left[ {-{{\left( {r{\
\partial _{r}}}\right) }^{2}}+{{\left( {\frac{{L_{z}}}{\hbar }}\right) }^{2}}
-2l\left( {\frac{{L_{z}}}{\hbar }}\right) {\sigma _{z}}}+l^{2}\right]
\nonumber \\
&&-\frac{{{\hbar ^{2}}\partial _{z}^{2}}}{{2M}}+\frac{{\tilde{\Omega}}}{2}
f(r){\sigma _{x}}+\frac{\tilde{\delta}}{2}{\sigma _{z}}+V(r),  \label{eq:H'0}
\end{eqnarray}
where $L_{z}=-i\hbar \partial _{\phi }$ is the $z$ component of
the angular momentum operator and $\{\sigma _{j}\}$ are Pauli
matrices. The SOAM coupling $L_{z}\sigma _{z}$ (as a part of more
general $\boldsymbol{L}\cdot { \ \boldsymbol{\sigma }}$ coupling)
emerges from such transformation, similar to the appearance of
$k_{x}\sigma _{z}$ in SLM coupling
experiments~\cite{Lin2011,Pan2012,Qu2013,Hamner2014,Olson2014,Ji2014,karina2014,Campbell2015,Wang2012,Cheuk2012,Williams2013}.

\section{Ring System}\label{sec:ring}

To reveal the most salient effects of SOAM
coupling in both theoretical and experimental aspects, we
investigate a ring BEC with a fixed radius $R$. Integrating out
the $z$ and $r$ dependence and using the natural energy unit
$\epsilon =\hbar ^{2}/(2MR^{2})$, we turn Eq.~(\ref{eq:H'0}) into
a dimensionless ring Hamiltonian,
\begin{eqnarray}
{H}_{0}^{\mathrm{ring}}=-\partial _{\phi }^{2}+\left( {2il{\partial _{\phi }}
+\frac{{\delta}}{2}}\right) {\sigma _{z}}+\frac{{\Omega}}{2}{\sigma _{x}},
\end{eqnarray}
where ${\delta}=\tilde \delta /\epsilon $ and ${\Omega}=\tilde
\Omega f(R)/\epsilon $ are the dimensionless detuning and Raman
coupling, respectively. Because
$[{{H}_{0}^{\mathrm{ring}},{L_{z}}}]=0$, the eigenstates of
${H}_{0}^{\mathrm{ring}}$ coincide with the OAM eigenstates $
|m\rangle $, or $e^{im\phi }$ with an integer $m$. The energy
spectrum shows two bands with the lowest one
\begin{eqnarray}
{E_{-}}(m)={m^{2}}-\frac{1}{2}\sqrt{{{\left( { \
4lm-{\delta}}\right) }^{2}}+{{\Omega}^{2}}}.
\end{eqnarray}
Applying the Hellmann-Feynman theorem, one can compute the spin
polarization from the energy spectrum as $\langle \sigma
_{z}\rangle =\partial E_{-}/\partial ({\delta}/2)$ and $\langle
\sigma _{x}\rangle =\partial E_{-}/\partial ({\Omega} /2)$.

For an assumed continuous spectrum, the ground state would
correspond to a real number $m^{\ast }$, analogous to the SLM
coupling case. In our system, however, due to OAM quantization,
the ground state does not exactly lie at $ m^{\ast }$ but the
nearest integer(s) $[m^{\ast }]$. Therefore, there can be two
degenerate ground states with adjacent quantum numbers
(reminiscent of a recently proposed idea of quantum time
crystal~\cite{Wilczek}). By letting $ E_{-}(m)=E_{-}(m+1)$ we
obtain a condition for degenerate $|m\rangle $ and $ |m+1\rangle$
as ${q_{m}}{\Omega}=\sqrt{(4{l^{2}} -q_{m}^{2})[{{(2l{q_{m}}-{\
\delta})}^{2}}-q_{m}^{2}]}$ with $q_{m}=2m+1$. If ${\delta}=0$,
the system has another twofold degeneracy $|\pm m\rangle $, except
for $m=0$. Combining these conditions, the non-interacting case
can exhibit at most fourfold degeneracy $|\pm m\rangle $ and $|\pm
(m+1)\rangle $. On the other hand, in the large ${\Omega}$ limit,
the system always has a single ground state $|0\rangle$. The
double minimum structure of $\left\vert \pm m\right\rangle$
degeneracy appears as ${\ \Omega}$ decreases across a critical
value ${\Omega}_{c}$, which can be evaluated as a threefold
degeneracy point of $|0\rangle $ and $|\pm 1\rangle $. We hence
obtain ${\Omega} _{c}=4l^{2}-1$. This is different from a
continuous spectrum because of the quantization of $m$. When
double minima appear at $\pm m^{\ast }$ closer to $ 0$ than $1$,
the system is enforced in the single state $|0\rangle $.

In Fig.~\ref{fig:f1}(b) we plot the ground-state phase diagram for
$l=2$. The OAM quantum numbers $m$ are labeled on the
corresponding blocks with borders in dashed lines, which also
represent regions with degeneracy. The spin polarization $\langle
\sigma _{z}\rangle $ displays discontinuity with the change of
$m$, and its sign is locked with the sign of $m$ for any non-zero
$m$. Both signatures can be directly attributed to the presence of
SOAM coupling. In Fig.~\ref{fig:f1}(c), we label the ground
state(s) on the assumed continuous spectrum at selected points
along the $ \delta=0$ line. We see the transition from
nondegenerate to various multidegenerate ground states as
${\Omega}$ varies. Remarkably, the threefold (curve ii) and
fourfold (iv) degeneracy does not occur in the continuous
spectrum.

\section{Interaction effects}\label{sec:interactioin}

We now analyze realistic systems with $s$-wave
scattering interactions. Incorporating the interactions
$g_{\uparrow }$ ($g_{\downarrow }$) between up (down) bosons and
the interspin bosons $g_{\updownarrow }$, the system's energy
reads as
\begin{eqnarray}
E=\int_{0}^{2\pi }{{\Psi ^{\dag }}\left( {{H} _{0}^{\mathrm{
ring}}+{H}_{g}^{\mathrm{ring}}}\right) \Psi d\phi },
\end{eqnarray}
where
\begin{eqnarray}
{H }_{g}^{ \mathrm{ring}}=\frac{1}{2}\left( {\
\begin{array}{cc}
{{g_{\uparrow }}{{\bar{\psi}}_{\uparrow }}{\psi _{\uparrow }}} & {{\
g_{\updownarrow }}{{\bar{\psi}}_{\downarrow }}{\psi _{\uparrow }}} \\
{{g_{\updownarrow }}{{\bar{\psi}}_{\uparrow }}{\psi _{\downarrow
}}} & {{\ g_{\downarrow }}{{\bar{\psi}}_{\downarrow }}{\psi
_{\downarrow }}}
\end{array}
}\right).
\end{eqnarray}
The normalization condition is set as
$\int_{0}^{2\pi }{{\Psi ^{\dag }}\Psi d\phi }=1$ such that
$g_{\uparrow ,\downarrow ,\updownarrow }$ are proportional to the
total number of particles $N$.

To capture the effects of SOAM coupling, interactions, and
possible degeneracies, we adopt a variational wave function of the
form
\begin{equation}
\Psi =\left( {{\Psi _{1}}+{e^{i\zeta }}{\Psi _{2}}}\right) /\sqrt{2\pi},
\end{equation}
where
\begin{eqnarray}
{\Psi _{j}}&=&\left\vert {C_{1}^{j}}\right\vert \left(
\begin{array}{c}
{\cos {\theta _{j}}} \\
{-\sin {\theta _{j}}}
\end{array}
\right) e^{i(m_{j}\phi +{\eta _{j}})} \nonumber\\
&&+\left\vert {C_{2}^{j}}\right\vert \left( {\
\begin{array}{c}
{\sin {\theta _{j}}} \\
{\ -\cos {\theta _{j}}}
\end{array}
}\right) {e^{-i(m_{j}\phi +{\eta _{j}})}} ,
\end{eqnarray}
with $m_{1}=m$ and $m_{2}=m+1$. The normalization condition gives
$\sum\nolimits_{i,j}{{{|{C_{i}^{j}}|}^{2}}} =1$. The range of
parameters is set to be $0\leq {\theta _{j}}\leq \pi /2$ and $-\pi
\leq {\eta _{j}},\zeta <\pi $. With this ansatz, we obtain $E$ as
a function of six independent parameters $|{C_{1}^{1}}|$,
$|{C_{1}^{2}}|$, $ | {C_{2}^{1}}|$, $\theta _{1}$, $\theta _{2}$,
and $\zeta $. The two phases $ \eta _{1}$ and $\eta _{2}$ do not
affect $E$ here but can play a role in a general case with
external potentials. These parameters are determined through the
minimization of $E$. In addition, we compare the variational
results with those from solving the Gross-Pitaevskii equation
(GPE) by the imaginary time evolution and find good agreement
between them.

With the interactions on, we obtain either $\Psi _{1}=0$ or $ \Psi
_{2}=0$, which indicates energetic disfavor of the superposition
of $ |m\rangle $ and $|m+1\rangle $. As a result, $\langle
|m|\rangle $ is always an integer and the phase $\zeta $ plays no
role. Below we assume $\Psi _{2}=0 $ for convenience.

\begin{figure}[t]
\centering
\includegraphics[width=8.6cm]{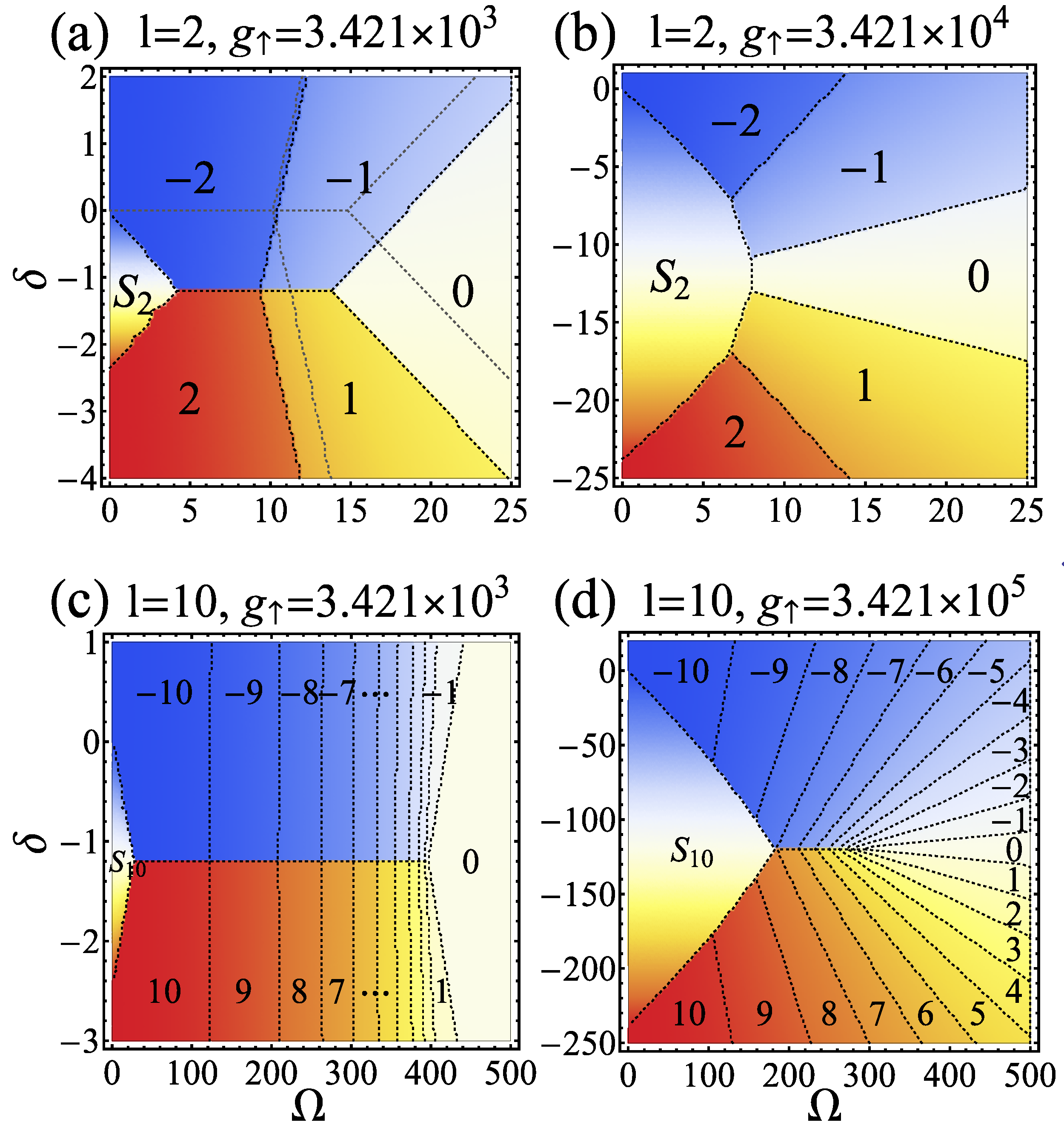}\vspace{-0.3cm}
\caption{(Color online) Phase diagrams with the presence of
interactions. (a) [(b)] corresponds to $l=2$ and $g_{\uparrow
}=3.421\times 10^{3}$ ($ 10^{4}$), and (c) [(d)] does to $l=10$
and $g_{\uparrow }=3.421\times 10^{3}$ ($ 10^{5}$). We set
$g_{\downarrow }=g_{\updownarrow }=0.9954g_{\uparrow}$, which is
good for $ ^{87}$Rb atoms. In (a) the noninteracting boundaries
are drawn in gray dashed lines for comparison. Conventions are the
same as Fig.~ \protect\ref{fig:f1}(b), except an emerging stripe
phase as a combination of $ |\pm m\rangle $ is denoted by $S_{m}$.
} \label{fig:f2}
\end{figure}

Figures \ref{fig:f2}(a) and \ref{fig:f2}(b) show phase diagrams
for $l=2$ at a fixed ratio $ g_{\downarrow }=g_{\updownarrow
}=0.9954g_{\uparrow }$. We present quantum numbers, phase
boundary, and spin polarization in the same convention as
Fig.~\ref{fig:f1}(b). The gray dashed curves in panel (a) show the
noninteracting phase boundary for comparison. We see that the
presence of interaction leads to (1) an emerging stripe phase and
(2) phase boundary shifts. In regions denoted with integer $m$,
the ground state lies in this specific quantum number, which means
only one of $|C_{1}^{1}|$ and $ |C_{2}^{1}|$ is non-zero, or
$|C_{1}^{1}C_{2}^{1}|=0$. Similar to the SLM coupling case, there
appears a region showing $|C_{1}^{1}C_{2}^{1}|\neq 0$,
corresponding to a linear combination of $|\pm m\rangle $ (denoted
by $S_{m}$ ). This state exhibits a spatial modulation in particle
density or a stripe structure since $\Psi ^{\dagger }\Psi
=1+2|C_{1}^{1}C_{2}^{1}|\sin 2\theta _{1}\cos 2(m\phi +\eta
_{1})$. The net spin polarization $\langle \sigma _{z}\rangle $ is
strongly suppressed in the stripe phase due to the cancellation
from $|\pm m\rangle $ with opposite polarizations. In contrast to
the SLM coupling case, the stripe phase here can still exhibit
significant spin polarization as a function of the detuning.

In panel (a), the vertical shifts of the phase boundary come from
the asymmetry of the interactions $g_{\uparrow }\neq g_{\downarrow
}$, which causes an effective Zeeman splitting $(g_{\uparrow
}-g_{\downarrow })/8\pi \times \langle \sigma _{z}\rangle $ in the
energy functional. This interaction-induced splitting, which
energetically favors down spins, competes with the detuning
${\delta }/2$ in its negative region. The phase boundary between
$|\pm m\rangle $ and the zero polarization line of the stripe
phase hence vertically shifts to a point ${\delta }\sim
-(g_{\uparrow }-g_{\downarrow })/4\pi $ where the two effects
balance. As $g_{\uparrow }$ increases by an order [from (a) to
(b)], the stripe phase $ S_{2}$ expands, invades the single-$m$
region, and finally intersects with all $m$ phases. At
intermediate stages, the boundary of $S_{2}$ can meet the point of
degenerate $|\pm 1\rangle ,|\pm 2\rangle $ to forms a fivefold
degeneracy and meet ${\Omega }_{c}$ (point of degenerate
$|0\rangle ,|\pm 1\rangle $) to form a fourfold degeneracy. We
notice that the $S_{1}$ phase is never energetically favorable
here. In addition, we find that ${\Omega } _{c}$ decreases with
the increase in $g_{\uparrow }$, indicating an interaction-induced
change between the single- and double-minimum
structures~\cite{Hamner2014,Li2012}.

\begin{figure}[t]
\centering
\includegraphics[width=8.6cm]{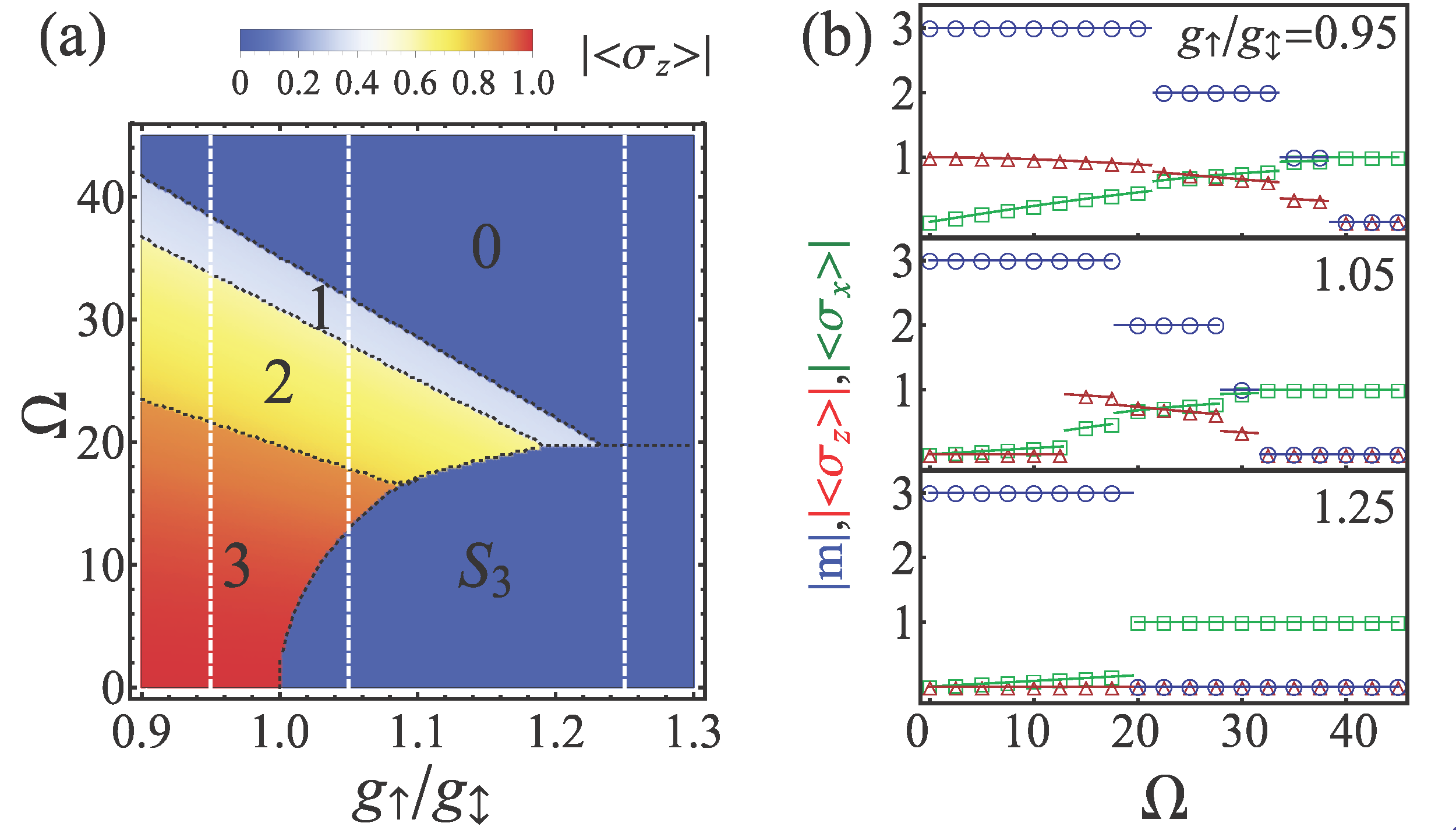}\vspace{-0.3cm}
\caption{(Color online) (a) Phase diagram in the $\Omega
$--$(g_{\uparrow }/g_{\updownarrow })$ plane. The $|m\rangle $
phases labeled by $|m|$ and the stripe phase $S_{3}$ are separated
by dashed lines. The colors represent spin polarization magnitude
$|\langle \protect\sigma _{z}\rangle |$ scaled in the bar graph.
The parameters are $l=3$, $\delta =0$, $ g_{\uparrow
}=g_{\downarrow }$, and $g_{\updownarrow }=424$ (good for $^{23}$
Na atoms). (b) From top to bottom: quantum number $|m|$ (circles)
and spin polarizations $|\langle \protect\sigma _{z}\rangle |$
(triangles) as well as $|\langle \protect\sigma _{x}\rangle |$
(squares) vs ${\Omega}$ at $ g_{\uparrow }/g_{\updownarrow
}=0.95$, $1.05$, and $1.25$, respectively, corresponding to the
white dashed lines in (a).} \label{fig:f3}
\end{figure}

For a larger $l$ case, the structure of the phase diagram remains
the same: the stripe phase on the left, $m\geq 1$ phases
decreasing from $|m|=l$ to $|m|=1$ in the middle, and $m=0$ phase
on the right. Panels (c) and (d) are phase diagrams for a case of
higher-order LG beams with $l=10$. In (c) we see the same
structure as the $l=2$ case in (a). The stripe phase $S_{10}$
appears on the left between $-2.4 < \delta < 0$, while the zero
momentum phase $|0\rangle$ appears on the right. In the middle
region, the finite quantum number phases $|m\rangle$ monotonically
decreases from $m=10$ to $ m=1 $ if $ \delta < -1.2$, while $m$
changes sign if $ \delta > -1.2$. The magnitude and sign of
$\langle \sigma_z \rangle$ behave in the same trend as $m$. In (d)
we show strongly interacting effects by increasing the interaction
strength by 100 times. One sees that the stripe-phase region
significantly expands, the boundaries of single-$m$ phases become
more inclined, and the zero-momentum-phase region shrinks. Such a
trend is similar to the $l=2$ case in (a) and (b).

We turn to study a case where the ratio of intraspin and interspin
interactions varies. Figure \ref{fig:f3}(a) shows a phase diagram
as a function of ${\Omega}$ and $g_{\uparrow }/g_{\updownarrow }$,
given $ g_{\uparrow }=g_{\downarrow }$, $g_{\updownarrow }=424$,
$l=3$, and ${\delta} =0$. We see that the stripe phase $S_{3}$
exists only when the ratio $ g_{\uparrow }/g_{\updownarrow }>1$
and $m\geq 1$ phases disappear at large ratio. In
Fig.~\ref{fig:f3}(b) we plot $|m|$, $|\langle \sigma _{z}\rangle
|$, and $|\langle \sigma _{x}\rangle | $ vs ${\Omega}$ at
$g_{\uparrow }/g_{\updownarrow }=0.95$ (no stripe phase), $1.05$
(all phases), and $1.25$ (no finite-$m$ phase), corresponding to
the white dashed lines from left to right in Fig.~\ref{fig:f3}(a),
respectively. We see that the system finally becomes fully
polarized in $\langle \sigma _{x}\rangle $ at large ${\Omega}$ .
The discrete jumps of $\langle \sigma _{x}\rangle =\partial
E/\partial { \Omega}$ indicate first-order phase transitions
between stripe and non-stripe phases as well as between different
$m$ phases.

\section{External potential} \label{sec:potential}

\begin{figure}[t]
\centering
\includegraphics[width=8.6cm]{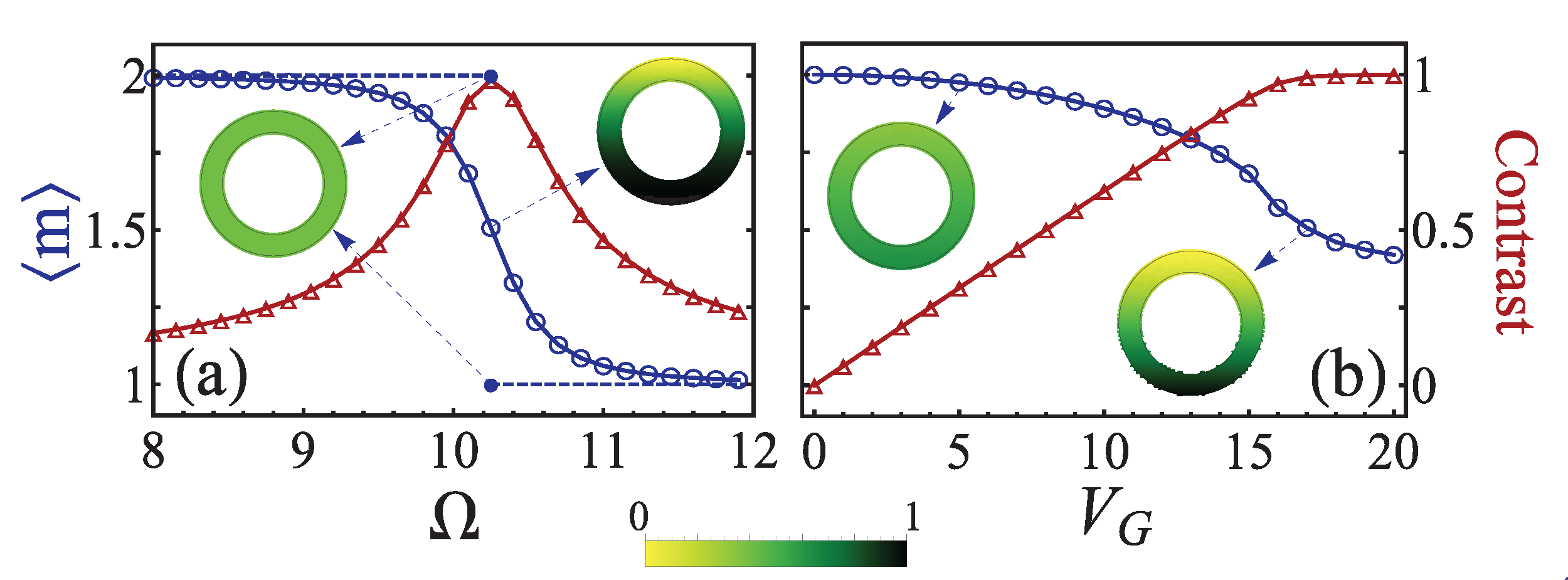}\vspace{-0.3cm}
\caption{(Color online) (a) Expectation value $\langle m \rangle$
(circles, axis on left) and density contrast [triangles, axis on
right of (b)] vs $ \Omega$ at gravity strength $V_G=0.05$ in a
noninteracting system. The dashed curve shows $\langle m \rangle$
at $V_G=0$ for comparison. Insets: normalized ring density
profiles (scaled in bar graph) for the two cases at $
\Omega=10.25$, indicated by the arrows, respectively. (b) Same
quantities vs $ V_G$ at $\Omega=10$ and
$g_\uparrow=g_\downarrow=g_\updownarrow=100$.} \label{fig:f4}
\end{figure}

We consider effects of a gravitational potential $
Ma_{g}R\sin \alpha \cos \phi $, where $a_{g}$ is the gravitational
acceleration and $\alpha $ is the angle between the ring plane and
the horizontal plane. For the mass of $^{23}$Na, $R=8$ $\mu $m, we
obtain a dimensionless gravity strength ${V}_{G}\equiv Ma_{g}R\sin
\alpha /\epsilon =1315\sin \alpha $, comparable to the interaction
strength as shown later. The gravity couples two adjacent OAM
states because $\left\langle m\right\vert \cos \phi \left\vert
{m\pm 1}\right\rangle =\frac{1}{2} \left\langle m\right\vert
{e^{i\phi }}+{e^{-i\phi }}\left\vert {m\pm 1} \right\rangle \neq
0$, so it should play a crucial role when the two states are
nearly degenerate. In such a case the variational ground state can
have both $\Psi _{1}$ and $\Psi _{2}$ non-zero. To pinpoint this
effect, we first study the transition region between $m=1$ and $2$
in the non-interacting case with tiny detuning [along the ${\delta
}=-0.01$ line in Fig.~\ref{fig:f1}(b)]. We plot $\langle m\rangle
$ and density contrast (defined as the normalized difference $
\frac{\rho _{M}-\rho _{m}}{\rho _{M}+\rho _{m}}$ between density
maximum $\rho _{M}$ and minimum $\rho _{m}$) vs ${\Omega }$ at
${V}_{G}=0.05$ in Fig.~\ref{fig:f4}(a). In contrast to the
discontinuity of the $V_{G}=0$ case (dashed curve), $\langle
m\rangle $ at $ V_{G}=0.05$ goes smoothly from 2 to 1, indicating
a mixed state around the transition point ${\Omega }=10.25$. Such
a state exhibits an inhomogeneous density profile (i.e., a stripe)
that is qualitatively different from the uniform one at $V_{G}=0$
(see inset). This makes the system a very sensitive detector for
gravity (${V}_{G}=0.05$ corresponds to $\alpha \lesssim 10^{-4}$
). Figure \ref{fig:f4}(b) shows the same quantities vs ${V}_{G}$
at ${\Omega } =10$, $g_{\uparrow }=g_{\downarrow }=g_{\updownarrow
}=100$, and $\delta =0$ , obtained from GPE. (The variational
results deviate at large $V_{G} $ due to the truncation of the
Hilbert space.) The contrast linearly increases with ${V}_{G}$ and
saturates when ${V}_{G}>16$. The sensitivity is hence controllable
through the tuning of $\Omega $ and the interactions.

Another experimentally feasible potential is an anisotropic trapping $\frac{1}{2}m{%
\omega ^2}[{x^2} + (1 - {\lambda ^2}){y^2}] = \frac{1}{4}m{\omega
^2}{R^2}{ \lambda ^2}\cos 2\phi $ up to a constant. This potential
couples $| m \rangle $ and $| {m \pm 2} \rangle$ and is expected
to stabilize the stripe phase composed of $|\pm 1\rangle$. The
gravity and anisotropic trapping are also capable of inducing
dipole and quadrupole density oscillations, respectively, for
studying the ring's collective excitations.

\section{Experimental aspects}\label{sec:experiment}

For a $^{87}$Rb gas trapped in a ring of radius
$R=20$ $\mu $m and thickness $b=5$ $\mu
$m~\cite{Ramanathan2011,Wright2013}, we have $\epsilon =2\pi \hbar
\times 0.145$ Hz. The dimensionless interaction strength can be
evaluated as $g=8NRa_{s}/b^{2}$ with the two-body scattering
length $a_{s}$~\cite{Olshanii,Bergeman}. The intra- and inter-spin
scattering lengths fix the ratio $g_{\downarrow }=g_{\updownarrow
}=0.9954g_{\uparrow }$~\cite{Lin2011}. For $a_{s}=100.86a_{0}$
(Bohr radius) and $N=10^{5}$, we obtain $g_{\uparrow }=3.421\times
10^{3}$ (as used in Fig.~\ref{fig:f2}). One can enhance $\epsilon
$ to $2\pi \hbar \times 0.91$ Hz by shrinking the ring size to
$R=8$ $\mu $ m, which, combined with higher-order LG beams of
$l=10$~\cite{Moulder2012}, gives $\Omega _{c}=2\pi \hbar \times
363$ Hz. For a $^{23}$Na gas~\cite{Stenger1998} with $R=8$ $\mu $m
and $l=10$, we get $\epsilon =2\pi \hbar \times 3.43$ Hz and
$\Omega _{c}=2\pi \hbar \times 1369$ Hz. Given $b=2$ $\mu $m,
$N=10^{4}$, and $ a_{s}=50$ $a_{0}$, typical interaction strength
is equal to $424$ $\epsilon $ (as used in Fig.~\ref{fig:f3}). For
typical $\Omega \simeq 1$ kHz, the heating effect due to
spontaneous photon emission of Raman lasers should be weak for a
typical experimental time scale of 1 s~\cite{Wei2013}. We notice
that, because $\epsilon $ can be so small, the interaction energy
[$ O(g/2\pi )$] can be much larger than the kinetic energy
[$O(l)$] and even $ \Omega _{c}$. Therefore, unlike the current
$^{87}$Rb platform where interactions show little competition with
the SLM coupling, our ring system is instead suited for exploring
the strongly interacting regime, where the ground-state phase
diagram could be significantly different from the noninteracting
case. For experimental detection, the quantum number $m$
corresponding to a superfluid winding number can be determined by
absorption images of the BEC after time-of-flight (TOF)
expansion~\cite{Wright2013}. The stripe phase will maintain its
pattern during TOF~\cite{Moulder2012}. Finally, we note that there
is ongoing experimental effort for generating such SOAM coupling
using $^{87}$Rb atoms confined on a ring trap~\cite{DAMOP} .

\section{Conclusion}\label{sec:conclusion}

A realistic scheme for generating SOAM coupling in
cold atom gases is proposed and analyzed. Study of the
ground-state phase diagram of the SOAM-coupled BEC on a ring
reveals the strong effects of many-body interaction with the
currently experimentally available parameters. The results should
provide a new platform for exploring SOAM-coupled cold atomic
physics for both bosons and fermions. Generalization of the scheme
for the full $\boldsymbol{L}\cdot {\ \boldsymbol{\sigma }}$
coupling may involve more LG laser beams and additional hyperfine
states, but may bring new exotic physics.

\section*{Acknowledgements}
We are grateful to L. Jiang, Y.-J. Lin, Y. Xu, and Z. Zheng for
interesting discussions. This work is supported by ARO
(W911NF-12-1-0334) and AFOSR (FA9550-11-1-0313 and
FA9550-13-1-0045). We acknowledge the Texas Advanced Computing
Center (TACC) for computational resources.

\end{document}